\documentclass[12pt]{article}
\usepackage{amsmath, amssymb}
\usepackage[dvips]{graphicx}
\usepackage{amsmath}
\newcommand{\spc}{\hspace{0.22in}}
\def \ds {\displaystyle}
\def \no {\noindent}
\def \it {\textit}

\def \lt {\(<\)}
\def \gt {\(>\)}

\def \es {\enspace}

\def \ol {\overline}
\title{
  \vspace*{20mm}
  {\Large {\bf Space-Time Duality \\
  and Vacuum Energy}} \\
  \vspace*{40mm}
}  
\author{
Hirotaka Sugawara
\thanks{E-mail address: ~sugawara\_hirotaka@soken.ac.jp, \es
{\it{Hayama Center for Advanced Studies, The Graduate University for Advanced Studies, 
Shonan International Village, Hayama, Kanagawa  240-0193  Japan, and Department of Physics and Astronomy, 
University of Hawaii at Manoa, Honolulu, HI  96822  U.S.A.}}}
}  
\date{}  
\begin{document}
\maketitle
\begin{center}
{\it{Hayama Center for Advanced Studies,}} \\
{\it{The Graduate Unviersity for Advanced Studies (Sokendai),}} \\
{\it{Shonan International Village, Hayama, Kanagawa  240-0193  Japan}} \\
and \\
{\it{Department of Physics and Astronomy,}} \\
{\it{University of Hawaii at Manoa, Honolulu, HI  96822  U.S.A.}}
\end{center}
\vspace{15mm}
\begin{abstract}
\vspace{5mm}

\spc
The concept of Euclidean time is proposed which is dual to the usual 
Minkowski time.
The De Sitter solution is shown to be dual to the anti-De Sitter solution
under the dual transformation in which Euclidean time and Minkowski 
time are interchanged.
This observation enables us to make a proposal in which the supersymmetry
is broken, but the four dimensional cosmological constant remains zero.
An explicit model is proposed in which we calculate the cosmological 
constant.
The phenomenology based on this model is presented, including the issue 
of the quark-lepton-Higgs mass matrix.
\end{abstract}
\vspace{15mm}
\newpage
%
%
\section{Introduction}\label{introduction} 
\vspace{10mm}

\spc
In 1999 a new type of solution to the five dimensional gravity equation 
was found by L. Randall and R. Sundrum~\cite{Randall_Sundrum}
.
However, its relation to other solutions, such as the ordinary De Sitter 
solution, or its implication to the cosmological constant 
problem~\cite{cosmological_constant} 
has not been understood.
Part of this work is devoted to clarifying this issue.
\\

Let us consider the Einstein equation with the dimension $N \geqq 6$ and 
with the cosmological constant $\lambda_{N}$ without matter.
The value of $\lambda_{N}$ can be positive, negative or zero.
We search for a solution in which the time direction ($x_{0}$) and one of 
the space directions in the extra dimension ($x_{5}$) (4 is reserved for 
$x_{4} = ix_{0}$) are treated differently from the other directions 
$x_{\mu} \, (\mu = 1, \, 2, \, 3)$ or $x_{i} \, (i \geqq 6)$.
Namely, we have a set of equations for the Ricci tensor components 
$R_{\mu \nu} \, (\mu, \, \nu = 0, \, 1, \, 2, \, 3)$,
$R_{\mu i} \, (\mu = 0, \, 1, \, 2, \, 3, \, i \geqq 5)$ and 
$R_{ij} \, (i, \, j \geqq 5)$ and we search for a solution where 
we can treat $x_{0}$ and $x_{5}$ differently from the others.
\\

What we find is the following:
\begin{itemize}
\item[(1)] $\lambda$ \, \gt \, $0$ ~~~~~usual expanding solution
\item[(2)] $\lambda$ \, \lt \, $0$ ~~~~~warp solution of Randall-Sundrum type \\
~~~~~~~~~~~~~~~(without $\delta$ function source in the equation)
\item[(3)] $\lambda \; = \; 0$ ~~~~~some dimensions expand, but other dimensions 
may expand, shrink \\
~~~~~~~~~~~~~~~or stay the same.
\end{itemize}
We also find that the solution of $\lambda$ \gt $\; 0$ is dual to the 
solution of $\lambda$ \lt $\; 0$ and so we can get one from the other 
by exchanging some variables.
In particular, Euclidean time and Minkowski time play a dual 
role in these solutions.
\\

In the absence of the $\delta$-function source term in the equation, 
the Randall-Sundrum solution can be interpreted as the five dimensional 
vacuum solution in which five dimensional vacuum energy exists, but the 
four dimensional spacetime is flat.
This gives a clue to understand why the large supersymmetry breaking 
may not result in the equally large vacuum energy:  five dimensional 
vacuum energy is large and negative, but the four dimensional spacetime 
is flat.
It is our view that this is the most important aspect of the Randall-Sundrum 
solution.

The geometry and the matter distribution must be self-consistent in a model 
where no arbitrary cosmological constant is allowed since the vacuum energy 
completely determines the cosmological constant and the latter determines the 
geometry of the vacuum at the classical level.

Supersymmetric models can guarantee the absence of the arbitrary 
cosmological constant and also the absence of vacuum energy.
The supersymmetry breaking implies the appearance of the vacuum energy.
The problem is to calculate it explicitly.
\\

As long as the energy is four dimensional, its value seems to be too large
to accomodate the experimental value of $\simeq 0.01 \; \rm{eV}$~\cite{WMAP_SDSS}.
By making the energy five dimensional, we can keep the 
four dimensional part flat, as the Randall-Sundrum solution suggests.
We construct an explicit model which accomodates this situation.

Following string theory or M-theory, we start from the pure $E(8)$ 
gauge theory, but the actual gauge symmetry we consider
is the $SU(3) \times E(6)$ which we obtain by breaking the $E(8)$ group in an
appropriate manner.
We assume the matter fields (including gauge and chiral fields) are localized
on a Euclidean time $\otimes$ four dimensional spacetime $M_{5}$.
Then the supersymmetry breaking gives rise to the five dimensional cosmological
constant.
The supergravity multiplet can lie on the bulk, but we assume there is no 
supersymmetry breaking in the bulk (massless gravitino).
The vacuum solution in the bulk, therefore, is given by the solution of the 
Einstein equation with $\lambda = 0$.
The bulk spacetime is given by $M_{4} \otimes C_{3}$ where $C_{3}$ is a 
cylinder with its axis as Euclidean time and $M_{4}$ is our usual spacetime.
We regard $M_{4} \otimes C_{3}$ is a fibre space in the M-theory bundle with
some Ricci flat K\"{o}hler manifold as the base.
The localization and the chiralization of some of our matter fields must be 
the consequence of some singularity in this manifold~\cite{Katz_et_al}.
\\

We address the question of how we obtain the finite value for the four 
dimensional cosmological constant which was observed by WMAP or SDSS.
The perturbative energy, such as the zero-point energy or its gauge 
corrections, may be of five dimensional character, but there will be an 
(vacuum) energy source which is purely of the four dimensional character.
The instanton contribution~\cite{Hooft} is an example.
We follow the work of Affleck, Dine and Seiberg~\cite{Affleck_Dine_Seiberg} 
on this issue and explore the possible consequence of its contribution 
in our model if the instanton actually contributes to the vacuum energy.

The interplay of the $SU(3)$ instanton and the $E(6)$ instanton leads to the
fixing of the flat direction by giving an increasing or a decreasing potential,
respectively.
The resulting cosmological constant is $v^{4} \, \exp \, ({-16 \pi^{2}} / {g^{2} (v)})$
with the $v$ of the Planck scale.
The value is amusingly consistent with the experimental value when we take 
$\alpha (v) = {{g^{2} (v)} \over 4 \pi}$ to be the unification
coupling constant which is slightly smaller than $1/20$~\cite{OPAL}.

Instanton or not, we assume that no superpotential is allowed unless it is 
dynamically generated, as in the cases treated by Affleck, Dine and Seiberg.
We have only three terms in our case of $SU(3) \times E(6)$
gauge model with $(3, \, 27)$ and $(\ol{3}, \, \ol{27})$ chiral multiplets 
if the renormalizability is imposed.
This gives us a means to determine the quark-lepton-Higgs mass directly in 
terms of Planck mass.
The essential mechanism is the fixation of the parameters of the flat direction.
\\

The organization of the article is the following: In section \ref{solution} 
we discuss the solutions of the $n$-dimensional vacuum Einstein equation 
with $\lambda$ \gt $\; 0$, $\lambda$ \lt $\; 0$ or $\lambda \, = \, 0$.
We also discuss the emerging geometrical structure of our vacuum spacetime 
based on these solutions.
In section \ref{possible model} we propose a model loosely based on 
string theory or M-theory.
The origin of the cosmological constant is the main theme of this section.
In section \ref{phenomenology} we discuss other phenomenological consequences 
of our model proposed in section \ref{possible model}.
The quark-lepton-Higgs mass is the main topic of this section.
Some additional comments will be presented in section \ref{conclusion}.

\vspace{15mm}
%
%
%
%
\section{Solution to $n$-Dimensional Einstein Equation}\label{solution}
\vspace{10mm}

\spc 
Solutions to the $n$-dimensional Einstein equation are obtained 
in this section for all the cases of 
$\lambda_{n}$ \gt $\; 0$, 
$\lambda_{n}$ \lt $\; 0$
and
$\lambda_{n} = \; 0$
where $\lambda_{n}$ is the $n$-dimensional cosmological constant.
It is shown that the 
$\lambda_{n}$ \gt $\; 0$ \,
solution is dual to the 
$\lambda_{n}$ \lt $\,\; 0$ \,
solution.
This is due to our ansatz, inspired by the Randall-Sundrum 
solution~\cite{Randall_Sundrum},
that there exists a fifth direction (with the Euclidean signature) 
which plays the role of compact Euclidean time and is dual 
to ordinary Minkowski time.
One of the main consequences of this section is the discovery of the 
non-trivial $\lambda_{n} = 0$ \, solution.

For $\lambda_{n} = 0$, we find a solution in which the usual three 
dimensional space is expanding, but the internal space may be shrinking.
This fact enables us to construct a rather realistic model in 
section \ref{possible model}.
The vanishing of the four dimensional vacuum energy is related to the 
$\lambda_{n}$ \lt $\; 0$ solution.
Our solution in this case is essentially that of Randall-Sundrum, 
although our equation has no $\delta$-function source term contrary 
to the Randall-Sundrum case.
Our $\lambda_{n}$ \lt $\; 0$ solution applied to five-dimensions indicates 
that four dimensional subspace is flat where the five dimensional 
cosmological constant $\lambda_{5}$ is negative.
This means that five dimensional supersymmetry can be broken with 
the negative $\lambda_{5}$,
but the corresponding four dimensional subspacetime is flat.
The entire matter must be such that it gives the negative $\lambda_{5}$.
Of course, we must start from a supersymmetric theory where no arbitrary 
cosmological constant is allowed.
All the sources of the cosmological constant must be attributable to 
vacuum energy.
\\

We start from the $n$-dimensional Einstein action without matter, but 
with the cosmological constant.
\begin{eqnarray} 
  {\cal L} = \frac{\ds -1}{\ds 2 \kappa^{2}_{N+5}} \int d^{N+5} \, x \, \sqrt{- \raisebox{1mm}{$g$}} \, [R + \lambda_{N+5}] \, . 
  \label{eq:2.1}
\end{eqnarray}
The metric is chosen to be 
$(- \, + \, + \, \cdots)$.  
R stands for the scalar curvature and 
$\lambda_{N+5}$ 
is the $N+5$ dimentional cosmological constant.

We put
$n = N+5$
to indicate that we are treating five dimensional spacetime with 
the metric $(- \, + \, + \, + \, +)$ differently from the rest.
The fifth coordinate, however, belongs to internal space and small 
Roman letters will be used for it just as other internal space
coordinates.
\\

The discussion will be general, but the main interest will be 
$N = 2$ case when we build a model which is based on M-theory 
or string theory.

From \eqref{eq:2.1}, we get the Einstein equation
\begin{equation} 
  R_{MN} = - {1 \over {N+3}} \, \lambda_{N+5} \, \raisebox{1mm}{$g$}{}_{MN} \, 
  \label{eq:2.2}
\end{equation}
\no
where $R_{MN}$ and $\raisebox{1mm}{$g$}_{MN}$ are ordinary Ricci 
and metric tensors.
\\

Let us use small Greek letters $\mu$, $\nu$, $\dots$ for 
$\mu, \, \nu, \, \rho, \, \dots \, = \, 0, \, 1, \, 2, \, 3$ and small 
Roman letters $i, \, j, \, k, \, l, \, m, \, n, \, \dots$ for 
$i, \, j, \, k, \, l, \, m, \, n, \, \dots \, = \, 5, \, 6, \, 7, \, 8, \, \dots$,  
to distinguish ordinary and internal spacetime coordinates.  
$i \, ($or $j, \, k, \, l, \, m, \, n, \, \dots) \, = \, 5$ plays 
a special role as Euclidean time in the internal space.

\no
The first ansatz we make is
\begin{gather*} 
  dS^{2} = \raisebox{1mm}{$g$}{}_{\mu \nu} \, dx^{\mu} dx^{\nu} 
    + \raisebox{1mm}{$g$}{}_{mn} \, dx^{m} dx^{n} \, ,
  \label{eq:2.3}
  \\
  i.e.~~%
  \raisebox{1mm}{$g$}{}_{\mu m} \; = \; 0 \, .
\end{gather*}

We also assume that $g_{\mu \nu}$ is a function of 
$x_{\rho} \, (\rho \, = \, 0, \, 1, \, 2, \, 3)$ 
and $x^{i}$, and $g_{mn}$ is a function of 
$x^{i} \, (i \, = \, 5, \, 6, \, 7, \, 8, \dots)$ 
and $x^{0} \, = \, t$:
\begin{gather*} 
  \raisebox{1mm}{$g$}{}_{\mu \nu} 
    = \raisebox{1mm}{$g$}{}_{\mu \nu} \, (x^{\rho}, \, x^{i}) \, ,
  \label{eq:2.4} 
  \\
  \raisebox{1mm}{$g$}{}_{mn} 
    = \raisebox{1mm}{$g$}{}_{mn} \, (x^{i}, \, x^{0} = t) \, . 
  \label{eq:2.5} 
\end{gather*}
\no
Eventually, we restrict the $x^{i}$ dependence of $g_{\mu \nu}$ 
to be only for $i \, = \, 5$ (Euclidean time).

From now on, we use $\xi^{i}$ rather than $x^{i}$, to clearly 
distinguish external and internal space parameters.
\\

After the standard manipulation we get
\begin{eqnarray} 
  R_{\mu \kappa} 
  &=& 
    \overline{R}_{\mu \kappa} 
    + \frac{\ds 1}{\ds 2} \raisebox{1mm}{$g$}{}^{ln} 
      \frac{\ds \partial^{2} g_{\mu \kappa}}{\ds \partial \xi^{n}{\partial \xi^{l}}} 
    + \raisebox{1mm}{$g$}{}^{\lambda \nu} \raisebox{1mm}{$g$}{}_{lm} \, 
      [\Gamma_{\nu \lambda}^{l} \Gamma_{\mu \kappa}^{m} 
    - \Gamma_{\kappa \lambda}^{l} \Gamma_{\mu \nu}^{m}] \nonumber \\
  &+&
    \raisebox{1mm}{$g$}{}^{ln} \, 
      [- \raisebox{1mm}{$g$}{}_{\eta \sigma} \Gamma_{\kappa l}^{\eta} \Gamma_{\mu \rm n}^\sigma 
    + \raisebox{1mm}{$g$}{}_{mp} \Gamma_{nl}^{m} \Gamma_{\mu \kappa}^{p}] \nonumber \\
  &+&
    \raisebox{1mm}{$g$}{}^{ln} \left[ \frac{\ds 1}{\ds 2} 
      \frac{\ds \partial^{2} g_{ln}}{\ds \partial x^{\mu} \partial x^{\kappa}} 
    + \raisebox{1mm}{$g$}{}_{\eta \sigma} \Gamma_{nl}^{\eta} \Gamma_{\mu \kappa}^{\sigma} 
    - \raisebox{1mm}{$g$}{}_{mp} \Gamma_{\kappa l}^{m} \Gamma_{\mu n}^{p} \right] \nonumber \\
  &=&
    - \frac{\ds 1}{\ds N+3} \lambda_{N+5} \, \raisebox{1mm}{$g$}{}_{\mu \kappa} \, , 
  \label{eq:2.6} 
\end{eqnarray}
\begin{eqnarray} 
  R_{mk} 
  &=& \overline{R}_{mk}
    + \frac{\ds 1}{\ds 2} \raisebox{1mm}{$g$}{}^{\lambda \nu} 
      \frac{\ds \partial^{2} g_{\lambda \nu}}{\ds \partial \xi^{m} \partial \xi^{k}} 
    + \raisebox{1mm}{$g$}{}^{\lambda \nu} \, 
      [- \raisebox{1mm}{$g$}{}_{\eta \sigma} \Gamma_{k \lambda}^{\eta} \Gamma_{m \nu}^{\sigma} 
    + \raisebox{1mm}{$g$}{}_{np} \Gamma_{\nu \lambda}^{n} \Gamma_{mk}^{p}] \nonumber \\
  &+&
    \raisebox{1mm}{$g$}{}^{ln} \raisebox{1mm}{$g$}{}_{00} \, [\Gamma_{nl}^{0} \Gamma_{mk}^{0} 
    - \Gamma_{kl}^{0} \Gamma_{mn}^{0}] \nonumber \\
  &+&
    \raisebox{1mm}{$g$}{}^{\lambda \nu} \left[ \frac{\ds 1}{\ds 2} 
      \frac{\ds \partial^{2} g_{mk}}{\ds \partial {x^ \lambda} \partial x^{\nu}} 
    + \raisebox{1mm}{$g$}{}_{00} \Gamma_{\nu \lambda}^{0} \Gamma_{mk}^{0} 
    - \raisebox{1mm}{$g$}{}_{np} \Gamma_{k \lambda}^{n} \Gamma_{m \nu}^{p} \right] \nonumber \\
  &=&
    - \frac{\ds 1}{\ds N+3} \lambda_{N+5} \, \raisebox{1mm}{$g$}{}_{mk}  \, , 
  \label{eq:2.7} 
\end{eqnarray}
\\
and
\begin{eqnarray} 
  R_{\mu k} 
  &=&
    \frac{\ds 1}{\ds 2} \raisebox{1mm}{$g$}{}^{\lambda \nu} 
      \left[ \frac{\ds \partial^{2} g_{\lambda \nu}}{\ds \partial x^{\mu} \partial \xi^{k}} 
    - \frac{\ds \partial^{2} g_{\mu \nu}}{\ds \partial x^{\lambda} \partial \xi^{k}} \right] \nonumber \\
  &+&
    \raisebox{1mm}{$g$}{}^{\lambda \nu} \raisebox{1mm}{$g$}{}_{\eta \sigma} \, 
      [\Gamma_{\nu \lambda}^{\eta} \Gamma_{\mu k}^{\sigma} 
    - \Gamma_{k \lambda}^{\eta} \Gamma_{\mu \nu}^{\sigma}] \nonumber \\
  &+&
    \raisebox{1mm}{$g$}{}^{\lambda \nu} \raisebox{1mm}{$g$}{}_{mp} \, 
      [\Gamma_{\mu k}^{p} \Gamma_{\nu \lambda}^{m} 
    - \Gamma_{k \lambda}^{m} \Gamma_{\mu \nu}^{p}] 
    + \frac{\ds \delta_{\mu}^{0}}{\ds 2} \raisebox{1mm}{$g$}{}^{ln} 
      \left[ \frac{\ds \partial^{2} g_{ln}}{\ds \partial x^{0} \partial \xi^{k}} 
    - \frac{\ds \partial^{2} g_{lk}}{\ds \partial x^{0} \partial \xi^{n}} \right] \nonumber \\
  &+&
    \delta_{\mu}^{0} \raisebox{1mm}{$g$}{}^{ln} \, 
      \{ \raisebox{1mm}{$g$}{}_{00} \, (\Gamma_{nl}^{0} \Gamma_{0k}^{0} 
    - \Gamma_{kl}^{0} \Gamma_{0n}^{0}) 
  + \raisebox{1mm}{$g$}{}_{ij} \, (\Gamma_{nl}^{i} \Gamma_{0k}^{j} 
    - \Gamma_{kl}^{i} \Gamma_{0n}^{j}) \} \nonumber \\
  &=& 
    0 \, .  
  \label{eq:2.8} 
\end{eqnarray}
Here, $\Gamma_{MN}^{L}$ is defined as usual:
\begin{eqnarray} 
  \Gamma_{MN}^{L} = \frac{\ds 1}{\ds 2} \raisebox{1mm}{$g$}{}^{LK} 
    \left[ \frac{\ds \partial g_{KN}}{\ds \partial x^{M}} 
    + \frac{\ds \partial g_{KM}}{\ds \partial x^{N}} 
    - \frac{\ds \partial g_{MN}}{\ds \partial x^{K}} \right] \, \raisebox{-4mm}{$.$}
  \label{eq:2.9} 
\end{eqnarray}
$\overline{R}_{\mu \kappa}$ is the Ricci tensor defined in terms of 
$g_{\mu \nu}$ as a function of $x^{\rho}$ and $\overline{R}_{mk}$ 
is defined in terms of $g_{kl}$ as a function of $\xi^{i}$.
\\

Now we proceed to search for a solution of the form:
\begin{equation} 
  \raisebox{1mm}{$g$}{}_{00} = - {W} \, (\xi^{i}, \, t) \, , 
  \label{eq:2.10}
\end{equation}
and
\begin{eqnarray} 
  \raisebox{1mm}{$g$}{}_{\mu \nu} 
    = R^{2} \, (\xi^{i}, \, t) \, \widetilde{\raisebox{1mm}{$g$}}_{\mu \nu} \, ,
  \, \; \mbox{with} \; \;
  \mu, \, \nu \, = \, 1, \, 2, \, 3 \; 
  \label{eq:2:11}
\end{eqnarray}
\no
with $\widetilde{g}_{\mu \nu}$ a three-\hspace{0pt}dimensional flat 
metric which depends only on $x^{\mu} \, (\mu \; = \; 1, \, 2, \, 3)$,
and
\begin{eqnarray} 
  \raisebox{1mm}{$g$}{}_{0 \mu} = 0 \quad \mbox{with} \quad \mu = 1, \, 2, \, 3 \, . 
  \label{eq:2.12}
\end{eqnarray}

This ansatz renders the equation for $R_{00}$ to be in the following form:
\begin{eqnarray} 
  \frac{\ds 3 \ddot{R}}{\ds R} - \frac{\ds 3}{\ds 2} \frac{\ds \dot{R}}{\ds R} \frac{\ds \dot{W}}{\ds W} 
  &+&
    \raisebox{1mm}{$g$}{}^{ln} \left[ - \frac{\ds 1}{\ds 2} \frac{\ds \partial^{2} W}{\ds \partial \xi^{l} \partial \xi^{n}} 
    - \frac{\ds 3}{\ds 4 R^{2}} \frac{\ds \partial R^{2}}{\ds \partial \xi^{l}} \frac{\ds \partial W}{\ds \partial \xi^{n}} 
    + \frac{\ds W}{\ds 4} \frac{\ds \partial \log W}{\partial \xi^{l}} \frac{\ds \partial \log W}{\ds \partial \xi^{n}} 
    + \frac{\ds 1}{\ds 2} \Gamma_{nl}^{j} \frac{\ds \partial W}{\ds \partial \xi^{j}} \right] \nonumber \\
  &+&
    \raisebox{1mm}{$g$}{}^{ln} \left[ \frac{\ds \partial^{2} g_{ln}}{\ds \partial t^{2}} 
    - \frac{\ds 1}{\ds 4} \frac{\ds \partial \log W}{\ds \partial t} \frac{\ds \partial g_{nl}}{\ds \partial t} 
    - \frac{\ds 1}{\ds 4} \raisebox{1mm}{$g$}{}^{mp} \frac{\ds \partial g_{ml}}{\ds \partial t} \frac{\ds \partial g_{pn}}{\ds \partial t} \right] \nonumber \\
  &=&
    \frac{\ds 1}{\ds N+3} \, \lambda_{N+5} \, W \, . 
  \label{eq:2.13}
\end{eqnarray}

\no
The equation for $R_{\mu \nu} \, (\mu, \, \nu \; =\;  1, \, 2, \, 3)$ becomes
\begin{eqnarray} 
  - \left(W^{-1} R \ddot{R} \right.
  &+&
    \left. 2 W^{-1} \dot{R}^{2} 
    + \frac{\ds W^{-2}}{\ds 2} R \dot{R} \dot{W} \right) \nonumber \\
  &+&
    \frac{\ds 1}{\ds 2} \raisebox{1mm}{$g$}{}^{ln} 
      \frac{\ds \partial R^{2}}{\ds \partial \xi^{l} \partial \xi^{n}} 
    + \frac{\ds 1}{\ds 2 R^{2}} \raisebox{1mm}{$g$}{}^{ln} 
      \frac{\ds \partial R^{2}}{\ds \partial \xi^{l}} 
      \frac{\ds \partial R^{2}}{\ds \partial \xi^{n}} \nonumber \\
  &+&
    \frac{1}{4} W^{-1} \raisebox{1mm}{$g$}{}^{ln} 
      \frac{\partial W}{\partial \xi^{l}} \frac{\partial R^{2}}{\partial \xi^{n}} 
    + \raisebox{1mm}{$g$}{}^{ln} \left[- \frac{\ds 1}{\ds 4 R^{2}} 
      \frac{\ds \partial \log R^{2}}{\ds \partial \xi^{l}} 
      \frac{\ds \partial \log R^{2}}{\ds \partial \xi^{n}} 
    - \frac{\ds 1}{\ds 2} \Gamma_{nl}^{m} 
      \frac{\ds \partial R^{2}} {\ds \partial \xi^{m}} \right] \nonumber \\
  &-&
    \frac{\ds 1}{\ds 2} W^{-1} R \dot{R} \, \raisebox{1mm}{$g$}{}^{ln} 
      \frac{\ds \partial g_{nl}}{\ds \partial t} \nonumber \\
  &=&
    - \frac{\ds 1}{\ds N+3} \, \lambda_{N+5} \, R^{2} \, 
  \label{eq:2.14} 
\end{eqnarray}
\no
and the equation for $R_{0 \mu} \, (\mu \, = \, 1, \, 2, \, 3)$ is 
trivially satisfied.
\\

The equation for $R_{mk}$ reads
\begin{eqnarray} 
  \overline{R}_{mk} 
  &+&
  \frac{\ds W^{-1}}{\ds 2} \frac{\ds \partial^{2} W}{\ds \partial \xi^{m} \partial \xi^{k}} 
    + \frac{\ds 3}{\ds 2 R^{2}} \frac{\ds \partial^{2} R^{2}}{\ds \partial \xi^{m} \partial \xi^{k}} 
    - 3 \left( \frac{\ds \partial \log R}{\ds \partial \xi^{m}} 
      \frac{\ds \partial \log R}{\ds \partial \xi^{k}} 
    + \frac{\ds 1}{\ds 2 R^{2}} 
      \frac{\ds \partial R^{2}}{\ds \partial \xi^{n}} \Gamma_{mk}^{n} \right) \nonumber \\
  &+&
    \left(- \frac{\ds 1}{\ds 4} \frac{\ds \partial \log W}{\ds \partial \xi^{k}} 
    \frac{\ds \partial \log W}{\ds \partial \xi^{m}} 
    - \frac{\ds 1}{\ds 2 W} \frac{\ds \partial W}{\ds \partial\xi^{n}} \Gamma_{mk}^{n} \right) 
  - \frac{\ds 1}{\ds 4 W} \raisebox{1mm}{$g$}{}^{ln} 
      \left( \frac{\ds \partial g_{nl}}{\ds \partial t} \frac{\ds \partial g_{mk}}{\ds \partial t} 
    - \frac{\ds \partial g_{kl}}{\ds \partial t} 
      \frac{\ds \partial g_{mn}}{\ds \partial t} \right) \nonumber \\
  &-&
    W^{-1} \left( \frac{\ds 1}{\ds 2} \frac{\ds \partial^{2} g_{mk}}{\ds \partial t^{2}} 
    - \frac{\ds W^{-1}}{\ds 4} \dot{W} \frac{\ds \partial g_{mk}}{\ds \partial t} 
    - \frac{\ds 1}{\ds 4} \raisebox{1mm}{$g$}{}^{ij} \frac{\ds \partial g_{ik}}{\ds \partial t} 
      \frac{\ds \partial g_{jm}}{\ds \partial t} \right) \nonumber \\
  &+&
    \frac{\ds 3}{\ds R^{2}} \left( - \frac{\ds 1}{\ds 2} R \dot{R} W^{-1} 
    \frac{\ds \partial g_{mk}}{\ds \partial t} \right) \nonumber \\
  &=&
    - \frac{\ds 1}{\ds N+3} \, \lambda_{N+5} \, \raisebox{1mm}{$g$}{}_{mk} \, .  
  \label{eq:2.15}
\end{eqnarray}

\no
The equation for $R_{\mu k}$ is trivially satisfied, except for the one 
for $R_{0k}$ which reads
\begin{eqnarray} 
  \frac{\ds 3 \dot{R}}{\ds 2 R} \frac{\ds \partial}{\ds \partial \xi^{k}} 
      \left( \log \frac{\ds \dot{R}^{2}}{\ds W} \right) 
  &-& \frac{\ds 3}{\ds 4 R^{2}} \raisebox{1mm}{$g$}{}^{nj} 
      \frac{\ds \partial g_{nk}}{\ds \partial t} \frac{\ds \partial R^{2}}{\ds \partial \xi^{j}} 
    + \frac{\ds 1}{\ds 2} \raisebox{1mm}{$g$}{}^{ln} \left( \frac{\partial^{2} g_{ln}}{\partial t \partial \xi^{k}} 
    - \frac{\ds \partial^{2} g_{lk}}{\ds \partial t \partial \xi^{n}} \right) \nonumber \\
  &-& \frac{\ds 1}{\ds 4} \raisebox{1mm}{$g$}{}^{ln} 
      \left( \frac{\ds \partial g_{nl}}{\ds \partial t} \frac{\ds \partial \log W}{\ds \partial \xi^{k}} 
    + \frac{\ds \partial g_{kl}}{\ds \partial t} \frac{\ds \partial \log W}{\ds \partial \xi^{n}} \right) \nonumber \\
  &=& 
    0 \, .  
  \label{eq:2.16}
\end{eqnarray}
Here, 
\begin{eqnarray*} 
  \dot{R} = \frac{\ds \partial R}{\ds \partial t} \, , \quad
  \ddot{R} = \frac{\ds \partial^{2} R}{\ds \partial t^{2}} \quad
  \mbox{and} \quad
  \dot{W} = \frac{\ds \partial W}{\ds \partial t} \, \raisebox{-3mm}{$.$}
\end{eqnarray*}
\\
So far we have not made any assumptions for the metric 
$g_{kl} \, (\xi, \, t)$.
\\

In order to obtain the solutions which are mutually dual when we exchange
$x^{0} \; = \; t$ and $\xi^{5}$ coordinates, we make the following ansatz 
for the internal space metric:
\begin{eqnarray} 
  d{S_{I}}^{2} = F (\xi^{5}, \, t) \, d{\xi^{5}} d{\xi^{5}} 
    + G (\xi^{5}, \, t) \textstyle\sum\limits_{l, \, n = 6}^{N+5} \delta_{ln} \, d \xi^{l} d \xi^{n} \, . 
  \label{eq:2.17}
\end{eqnarray}
\no
Here the fifth coordinate plays the role of Euclidean time 
among the internal space coordinates.
$I$ in $dS_{I}$ stands for the internal space. 
We also restrict the dependence of
$g_{\mu \nu} \, (\xi^{i}, \, x^{\rho})$ to be only on $\xi^{5}$ 
and $x^{\rho}$.
\\

With this ansatz the equation for $R_{00}$ becomes (using the notation 
$\xi \equiv \xi^{5}$):
\begin{eqnarray} 
  \frac{\ds 3 \ddot{R}}{\ds R} 
  &-&
    \frac{\ds 3}{\ds 2} \frac{\ds \dot{R}}{\ds R} \frac{\ds \dot{W}}{\ds W} 
    + F^{-1} \Bigl\{ - \frac{\ds 1}{\ds 2} \frac{\ds \partial^{2} W}{\ds \partial \xi^{2}}
    - \frac{\ds 3}{\ds 4 R^{2}} \frac{\ds \partial R^{2}}{\ds \partial \xi} \frac{\ds \partial W}{\ds \partial \xi} 
    + \frac{\ds W}{\ds 4} \left( \frac{\ds \partial \log W}{\ds \partial \xi} \right)^{2} 
    + \frac{\ds F^{-1}}{\ds 4} \frac{\ds \partial F}{\ds \partial \xi} 
      \frac{\ds \partial W}{\ds \partial \xi} \Bigr\}  \nonumber \\
  &-&
    \frac{\ds N}{\ds 4} G^{-1} F^{-1} \frac{\ds \partial W}{\ds \partial \xi} \frac{\ds \partial G}{\ds \partial \xi} 
    + F^{-1} \Bigl\{ \frac{\ds 1}{\ds 2} \frac{\ds \partial^{2}F}{\ds \partial t^{2}}
    - \frac{\ds 1}{\ds 4} \frac{\ds \partial \log W}{\ds \partial t} \frac{\ds \partial F}{\ds \partial t} 
    - \frac{\ds 1}{\ds 4} F^{-1} \left( \frac{\ds \partial F}{\ds \partial t} \right)^{2} \Bigr\}  \nonumber \\
  &+&
    NG^{-1} \Bigl\{ \frac{\ds 1}{\ds 2} \frac{\ds \partial^{2} G}{\ds \partial t^{2}} 
    - \frac{\ds 1}{\ds 4} \frac{\ds \partial \log W}{\ds \partial t} \frac{\ds \partial G}{\ds \partial t} 
    - \frac{\ds G^{-1}}{\ds 4} {\left( \frac{\ds \partial G}{\ds \partial t} \right)}^{2} \Bigr\} \nonumber \\
  &=&
    \frac{\ds 1}{\ds N+3} \, \lambda_{N+5} \, W \, , 
  \label{eq:2.18} 
\end{eqnarray}
\\
and the equation for $R_{\mu \nu} \, (\mu, \, \nu = 1, \, 2, \, 3)$ is:
\begin{eqnarray} 
  - \left( W^{-1} R \ddot{R} \right.
  &+&
  \left. 2 W^{-1} \dot{R}^2 - \frac{\ds W^{-2}}{\ds 2} R \dot{R} \dot{W} \right) \nonumber \\
  &+&
  F^{-1} \Bigl\{ R \frac{\ds \partial^{2} R}{\ds \partial \xi^{2}} 
    + 2 { \left( \frac{\ds \partial R}{\ds \partial \xi} \right)}^{2}
    + \frac{\ds R}{\ds 2} \frac{\ds \partial R}{\ds \partial \xi} W^{-1} \frac{\ds \partial W}{\ds \partial \xi}
    - \frac{\ds 1}{\ds 2} R \frac{\ds \partial R}{\ds \partial \xi} F^{-1} \frac{\ds \partial F}{\ds \partial \xi}
    - \frac{\ds 1}{\ds 2} W^{-1} R \dot{R} \frac{\ds \partial F}{\ds \partial t} \Bigr\} \nonumber \\
  &+&
  G^{-1} \Bigl\{ {-R \frac{\ds \partial R}{\ds \partial \xi} 
      \left( \frac{\ds {-NF}^{-1}}{\ds 2} \right) \frac{\ds \partial G}{\ds \partial \xi} 
    - \frac{\ds N}{\ds 2} W^{-1} R \dot{R} \frac{\ds \partial G}{\ds \partial t}} \Bigr\} \nonumber \\
  &=&
  - \frac{\ds 1}{\ds N+3} \, \lambda_{N+5} \, R^{2} \, . 
  \label{eq:2.19} 
\end{eqnarray}

Out of the equations for $R_{mk}$, $R_{5k}$ is trivially satisfied and the
$R_{55}$ equation becomes:
\begin{eqnarray} 
  \frac{\ds N}{\ds 2} G^{-1} \frac{\ds \partial^{2}G}{\ds \partial \xi^{2}} 
    &-&
    \frac{\ds N}{\ds 4} \left( \frac{\ds \partial \log G}{\ds \partial \xi} \right)^{2} 
    - \frac{\ds N}{\ds 4} \frac{\ds \partial \log F}{\ds \partial \xi} \frac{\ds \partial \log G}{\ds \partial \xi} \nonumber \\
  &+&
    \frac{\ds W^{-1}}{\ds 2} \frac{\ds \partial^{2} W}{\ds \partial \xi^{2}} 
    + \frac{\ds 3}{\ds 2 R^{2}} \frac{\ds \partial^{2} R^{2}}{\ds \partial \xi^{2}} 
    - 3 \Bigl\{ \left( \frac{\ds \partial \log R}{\ds \partial \xi} \right)^{2} 
    + \frac{\ds 1}{\ds 2 R^{2}} \frac{\ds \partial R^{2}}{\ds \partial \xi} 
      \left( \frac{\ds F^{-1}}{\ds 2} \frac{\ds \partial F}{\ds \partial \xi} \right) \Bigr\} \nonumber \\
  &+&
    \Bigl\{ - \frac{\ds 1}{\ds 4} \left( \frac{\ds \partial \log W}{\ds \partial \xi} \right)^{2} 
    - \frac{\ds W^{-1}}{\ds 2} \frac{\ds \partial W}{\ds \partial \xi} 
      \left( \frac{\ds F^{-1}}{\ds 2} \frac{\ds \partial F}{\ds \partial \xi} \right) \Bigr\} 
    - \frac{\ds N}{\ds 4} W^{-1} G^{-1} \frac{\ds \partial G}{\ds \partial t} \frac{\ds \partial F}{\ds \partial t} \nonumber \\
  &-&
    W^{-1} \Bigl\{ \frac{\ds 1}{\ds 2} \frac{\ds \partial^{2} F}{\ds \partial t^{2}} 
    - \frac{\ds W^{-1}}{\ds 4} \dot{W} \frac{\ds \partial F}{\ds \partial t} 
    - \frac{\ds 1}{\ds 4} F^{-1} \left( \frac{\ds \partial F}{\ds \partial t} \right)^{2} \Bigr\}  
    - \frac{\ds 3}{\ds 2 R} \dot{R} W^{-1} \frac{\ds \partial F}{\ds \partial t}  \nonumber \\
  &=& - \frac{\ds 1}{\ds N+3} \, \lambda_{N+5} \, F \, .  
  \label{eq:2.20} 
\end{eqnarray}

\no
$R_{mk}$ equation for $m, \, k = 6, \, \dots, \, N+5$ is:
\begin{eqnarray} 
  \frac{\ds \partial}{\ds \partial \xi} \left( \frac{\ds F^{-1}}{\ds 2} \frac{\ds \partial G}{\ds \partial \xi} \right)
    &-&
    \frac{\ds 1}{\ds 2} F^{-1} G^{-1} \left( \frac{\ds \partial G}{\ds \partial \xi} \right)^{2}
    + \frac{\ds F^{-1}}{\ds 4} \frac{\ds \partial G}{\ds \partial \xi} \left( NG^{-1} \frac{\ds \partial G}{\ds \partial \xi} 
    + F^{-1} \frac{\ds \partial F}{\ds \partial \xi} \right) \nonumber \\
  &+&
    \frac{\ds 3}{\ds 4} F^{-1} R^{-2} \frac{\ds \partial R^{2}}{\ds \partial \xi} \frac{\ds \partial G}{\ds \partial \xi}
    + \frac{\ds W^{-1} F^{-1}}{\ds 4} \frac{\ds \partial W}{\ds \partial \xi} \frac{\ds \partial G}{\ds \partial \xi} \nonumber \\
  &-&
    \frac{\ds N-2}{\ds 4} W^{-1} G^{-1} \left( \frac{\ds \partial G}{\ds \partial t} \right)^{2}
    - W^{-1} \left( \frac{\ds 1}{\ds 2} \frac{\ds \partial^{2} G}{\ds \partial t^{2}} 
    - \frac{\ds W^{-1}}{\ds 4} \dot{W} \frac{\ds \partial G}{\ds \partial t} \right) \nonumber \\
  &-&
    \frac{\ds 3}{\ds 2} R^{-1} \dot{R} W^{-1} \frac{\ds \partial G}{\ds \partial t}
    - \frac{\ds 1}{\ds 4} W^{-1} F^{-1} \frac{\ds \partial F}{\ds \partial t} \frac{\ds \partial G}{\ds \partial t} \nonumber \\
  &=& - \frac{\ds 1}{\ds N+3} \, \lambda_{N+5} \, G \, .  
  \label{eq:2.21}
\end{eqnarray}

Finally, the only non-trivial equation out of the equations for $R_{0k}$ 
(equation \eqref{eq:2.16}
) is for $R_{05}$ which reads:
\begin{eqnarray} 
  \frac{\ds 3}{\ds 2} R^{-2} \frac{\ds \partial^{2} R^{2}}{\ds \partial t \partial \xi} 
  &-&
    \frac{\ds 3}{\ds 4} \frac{\ds \partial \log R^{2}}{\ds \partial t} \frac{\ds \partial \log R^{2}}{\ds \partial \xi}
    - \frac{\ds 3}{\ds 4} \frac{\ds \partial \log R^{2}}{\ds \partial t} \frac{\ds \partial \log W}{\ds \partial \xi} \nonumber \\
  &+&
    \frac{\ds N}{\ds 2} G^{-1}\frac{\ds \partial^{2}G}{\ds \partial t \partial \xi}
    - \frac{\ds N}{\ds 4} \frac{\ds \partial \log G}{\ds \partial \xi} \frac{\ds \partial \log G}{\ds \partial t}
    - \frac{\ds N}{\ds 4} \frac{\ds \partial \log F}{\ds \partial t} \frac{\ds \partial \log G}{\ds \partial \xi} \nonumber \\
  &-&
    \frac{\ds 3}{\ds 4} \frac{\ds \partial \log R^{2}}{\ds \partial \xi} \frac{\ds \partial \log F}{\ds \partial t}
    - \frac{\ds N}{\ds 4} \frac{\ds \partial \log G}{\ds \partial t} \frac{\ds \partial \log W}{\ds \partial \xi} \nonumber \\
  &=&
    0 \, .  
  \label{eq:2.22}
\end{eqnarray}

We can now easily check that the $R_{00}$ equation is equivalent 
to the $R_{55}$ equation under the following changes:
\enlargethispage{7mm}
\begin{eqnarray} 
  t \; \longleftrightarrow \; \xi, \quad
  F \; \longleftrightarrow \; W, \quad
  G \; \longleftrightarrow \; R^{2} \quad \mbox{and} \quad
  \lambda_{(N+3)+2} \; \longleftrightarrow \; {- \lambda_{(N+3)+2}}  
  \label{eq:2.23}
\end{eqnarray}
\no
with $N \leftrightarrow 3$ in the coefficients of each term.
\\

This is the consequence of the ansatz made in equation \eqref{eq:2.17}
.  
We note the change of the sign of the cosmological constant, implying 
that we get the anti-De Sitter solution by the transformation of 
equation \eqref{eq:2.23}
\, (which may be called the space-time dual transformation) from 
the De Sitter solution.

Likewise, the $R_{\mu \nu}$ equation and the $R_{mk}$ equation are 
dual with respect to each other.  
The $R_{05}$ equation is self-dual under the transformations \eqref{eq:2.23}
.
\\

Let us discuss the three cases, $(1) \; \lambda$ \gt $\; 0$, 
$(2) \; \lambda$ \lt $\; 0$ and $(3) \; \lambda = 0$, separately:
\vspace{3mm}

\begin{itemize}
\item[(1)] \; $\; \lambda$ \gt $\; 0$
\\
\no
In this case we have an expanding solution of the form
\begin{gather} 
  W = W_{0} \, 
  \mbox{(constant),} \quad
  F = F_{0} \, e^{\beta t} \, , 
  \nonumber \\
  R^{2} = {R_{0}}^{2} \, e^{ct} 
  \quad \mbox{and} \quad
  G = G_{0} \, e^{\delta t} \, .  
  \label{eq:2.24}
\end{gather} 

We obtain the following equations from $R_{00}$, $R_{55}$, $R_{\mu \nu}$ 
and $R_{mk}$ equations respectively,
\begin{align} 
  \frac{\ds 3}{\ds 4} c^{2} + \frac{\ds 1}{\ds 4} \beta^{2} + \frac{\ds N}{\ds 4} \delta^{2} 
    &= \frac{\ds 1}{\ds N+3} \, \lambda_{N+5} W_{0} \, ,  
  \label{eq:2.25} \\
  \frac{\ds 1}{\ds 4} \beta^{2} + \frac{\ds N}{\ds 4} \delta \beta + \frac{\ds 3}{\ds 4} c \beta 
    &= \frac{\ds 1}{\ds N+3} \, \lambda_{N+5} W_{0} \, ,  
  \label{eq:2.26} \\
  \frac{\ds 3}{\ds 4} c^{2} + \frac{\ds 1}{\ds 4} c \beta + \frac{\ds N}{\ds 4} c \delta 
    &= \frac{\ds 1}{\ds N+3} \, \lambda_{N+5} W_{0} \, ,  
  \label{eq:2.27} \\
  \intertext{and}
  \frac{\ds N}{\ds 4} \delta^{2} + \frac{\ds 3}{\ds 4} c \delta + \frac{\ds 1}{\ds 4} \beta \delta 
    &= \frac{\ds 1}{\ds N+3} \, \lambda_{N+5} W_{0} \, .  
  \label{eq:2.28}
\end{align}
\no
$R_{05}$ equation is trivially satisfied.  
\\

\no
From equation \eqref{eq:2.25}
, we see that $\lambda_{N+5} W_{0}$ \gt $\; 0$ which means 
$\lambda_{N+5}$ \gt $\; 0$ $\left( W_{0} \right.$ \lt $\; 0$ 
is just the redefinition of time direction.  
We can always take $W_{0}$ \gt $\, \left. 0 \right)$.  
An obvious solution to \eqref{eq:2.25}
, \eqref{eq:2.26}
, \eqref{eq:2.27}
\, and \eqref{eq:2.28} 
is $c = \beta = \delta$ and
\[ \frac{\ds N+4}{\ds 4} \, c^{2} = \frac{\ds 1}{\ds N+3} \, \lambda_{N+5} \, W_{0} \, . \]
\item[(2)] \; $\lambda$ \lt $\; 0$
\\

\no
We get the following warp solution:
\begin{gather} 
  F = F_{0} \,
  \mbox{(constant),} \quad
  W = W_{0} \, e^{b \xi} \, ,  \nonumber \\
  G = G_{0} \, e^{\gamma \xi}
  \quad \mbox{and} \quad
  R^{2} = {R_{0}}^{2} \, e^{d \xi} \, . 
  \label{eq:2.285}
\end{gather}

\no
In fact, this can be obtained from the $\lambda$ \gt $\; 0$ solution by the 
dual transformation \eqref{eq:2.23} with the following parameter change:
\begin{eqnarray*} 
  (c, \, \beta, \, \delta, \, W_{0}, \, R_{0}^{2}, \, \lambda, \, 3, \, N) 
  \quad
  \Longleftrightarrow
  \quad
  (\raisebox{1mm}{$\gamma$}, \, b, \, d, \, F_{0}, \, G_{0}, \, -\lambda, \, N, \, 3) \, .
\end{eqnarray*}

\no
What we have achieved here is the proof that the expanding solution in the 
De Sitter case 
$\left( \lambda \right.$ \gt $\; \left. 0 \right)$ 
is dual to the warp solution in the anti-De Sitter case 
$\left( \lambda \right.$ \lt $\; \left. 0 \right)$.
\\

\no
The most important property of this warp solution is the flatness of four 
dimensional spacetime.
This is the property shared by the Randall-Sundrum solution which corresponds 
to the five dimensional 
Einstein equation with the $\delta$-function source term. 
\\

\no
We will be utilizing this property in model building where supersymmetry 
is broken in the five-dimensional theory, but leaves the four dimensional 
spacetime part flat.
The bulk space in our model admits only the supergravity multiplet and no 
supersymmetry is broken leading to $\lambda = 0$ in the bulk.
\\

\no
The vacuum solution in the bulk is then given by the solution with 
$\lambda = 0$.
No $\delta$-function source term will be created from the matter in the five 
dimensional subspace where the matter exists
because we impose the equation of motion, unlike in the case of Randall-Sundrum, 
both in the bulk and in the five dimensional subspace with matter.
\\
\item[(3)] \; $\lambda = 0$
\\
\no
Let us consider what happens to the $\lambda = 0$ case which we assume 
to happen in the bulk. 
\\

\no
One might suspect that we will get only a trivial solution (constant) 
in this case.  
Instead, we have the following form of non-trivial solution:
\begin{gather} 
  F = F_{0} \, e^{at + b \xi} \, , \quad
  W = W_{0} \, e^{at + b \xi} \, , \nonumber \\
  R^{2} = R_{0}^{2} \, e^{ct + d \xi} 
  \quad \mbox{and} \quad
  G = G_{0} \, e^{\delta t + \gamma \xi} \, .  
  \label{eq:2.29}
\end{gather}

\no
We get the following equations from $R_{00}$, $R_{55}$, $R_{\mu \nu}$ 
and $R_{mk}$ equations respectively:
\begin{gather} 
  \frac{\ds 3}{\ds 4} \, c \, (c - a) + \frac{\ds N}{\ds 4} \, \delta \, (\delta - a) 
    = \frac{\ds F_{0}^{-1} W_{0}}{\ds 4} \, b \, (3d + N \raisebox{1mm}{$\gamma$}) \, ,  
  \label{eq.2.30}
  \\
  \frac{\ds N}{\ds 4} \, \raisebox{1mm}{$\gamma$} \, (\raisebox{1mm}{$\gamma$} - b) 
    + \frac{\ds 3}{\ds 4} \, d \, (d - b) 
    = \frac{\ds F_{0} W_{0}^{-1}}{\ds 4} \, a \, (N \delta + 3c) \, ,  
  \label{eq.2.31}
  \\
  \frac{\ds c}{\ds 4} \, (3c + N \delta) 
    = \frac{\ds F_{0}^{-1} W_{0}}{\ds 4} \, d \, (3d + N \raisebox{1mm}{$\gamma$}) \, ,  
  \label{eq.2.32}
  \\
  \frac{\gamma}{4} \, (N \raisebox{1mm}{$\gamma$} + 3d) 
    =
    \frac{\ds F_{0} W_{0}^{-1}}{\ds 4} \, \delta \, (N \delta + 3c) \, .  
  \label{eq:2.33}
\end{gather}
\no
We can check that $R_{05}$ equation gives an identity.  
\\
\end{itemize}

In terms of $v^{2} = F_{0}^{-1} W_{0}$ we get
\begin{equation} 
  \delta = v \raisebox{1mm}{$\gamma$}
  \quad \mbox{and} \quad
  c = vd \, .  
  \label{eq.2.34}
\end{equation}
This means
\begin{equation} 
  R^{2} = R_{0}^{2} \, e^{c \left( t + {\xi}/{v} \right) }
  \quad \mbox{and} \quad
  G = G_{0} \, e^{ \delta \left( t + {\xi}/{v} \right) } \, .  
  \label{eq:2.35}
\end{equation}
The relation between $c$ and $\delta$ can be given by
\begin{equation} 
  {N \delta}^{2} - N \overline{a} \delta + 3c (c - \overline{a}) = 0 \,  
  \label{eq.2:37}
\end{equation}
where $\overline{a} = a + vb$.
\\

If we assume $c$ \gt $\; 0$ (expanding universe), we get either 
$\delta$ \gt $\; 0$ or $\delta$ \lt $\; 0$.  
$\delta = 0$ is realized when $c = \overline{a}$.  
The internal space, therefore, can expand, shrink or stay in the same size.  
If $c = 0$ (Minkowski case), $\delta$ is either positive or 0.  
\\

\begin{figure}[h] 
  \begin{center}
  \includegraphics[width=5cm, clip]{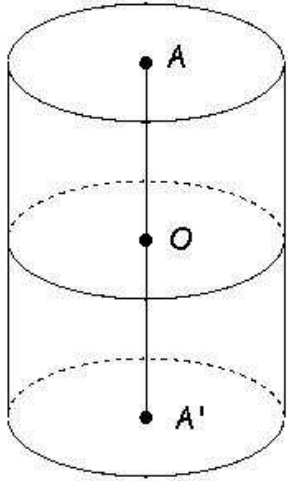}
  \end{center}
\vspace{-8mm}
\caption{A cylinder $C_{3}$ as the internal space}
\label{fig:1}
\vspace{3mm}
\end{figure}

The following image for our vacuum geometry emerges.  
Let us consider, as an example, the internal space to be a cylinder $C$ 
shown in fig.\,\ref{fig:1},
with $A$ and $A^{\prime}$ identified (whole bottom is identified with the top).

A four dimensional spacetime $M_{4}$ is attached to every point in the cylinder.  
Our total spacetime is $M_{4} \times C_{3}$.  

As in the case of supersymmetric deconstruction theory~\cite{Hamed_Cohen_Georgi}, 
we associate gauge multiplets and some of the chiral multiplets 
on the one-dimensional line $AOA^\prime$.  
Vacuum energy will exist along this line, due to these multiplets 
and the resulting breaking of the supersymmetry.
\\

Let us assume this vacuum energy to be negative.  
Then the vacuum solution on $M_{4} \times AOA^\prime$ is given by 
the five dimensional warp solution, and the corresponding four 
dimensional spacetime $M_{4}$ will be flat, no matter how we break 
the supersymmetry, as long as the five dimensional vacuum energy is negative. 

The bulk space has nothing but supergravity.
The vacuum solution in the $M_{4} \times (\rm{cylinder} - \rm{line} AOA^\prime)$ 
will be given by our $n$-dimensional vacuum solution with $\lambda \; = \; 0$ 
if the supergravity sector remains supersymmetric.  
%
%
\vspace{15mm}
%
%
%
\section{A Possible Model which is Consistent \\ 
with the Solution of the Previous Section}\label{possible model}
\vspace{10mm}

\spc
We describe a model in this section which is implied by our $n$-%
dimensional gravitational solutions of the last section.
If we start from M-theory and compactify the four dimensional part 
as a manifold ($K$) of Ricci flat and K\"{o}hler, the rest of the seven
dimensional spacetime will be regarded as fibre space with $K$ 
as the base space~\cite{Joyce}.

There is a possibility that some of the chiral multiplets are 
localized in the fibre space, depending on the singularity of the 
space $K$.
We assume that this fibre space is $M_4 \times C_3$ where $M_4$ is 
the four dimensional spacetime and that localization leads to the 
$M_{4} \times AOA^{\prime}$ as the space where matter can exist.
\\

The $AOA^{\prime}$ axis of the cylinder $C_{3}$ admits the supersymmetry 
breaking naturally because the compact space with a certain discrete 
symmetry is known to lead to these breakings.
In the case of supersymmetry it is the Sherk-Shwartz 
mechanism~\cite{Sherk_Schwarz} (although they considered $U(1)$ rather 
than discrete symmetry).
It is the boundary condition mechanism~\cite{boundary_condition_mechanism} 
in the case of gauge symmetry.
Our discrete symmetry is just the periodicity $(A \simeq A^{\prime})$.
\\

A crucial observation is that most of the supersymmetry breaking in 
five dimensional spacetime lead to the five dimensional vacuum energy, 
rather than the four dimensional one. 
We have 
$\lambda \, g_{\mu \nu} \, (\mu, \, \nu = 0, \, 1, \, 2, \, 3, \, 5)$ \, ,
rather than
$\lambda \, g_{\mu \nu} \, (\mu, \, \nu = 0, \, 1, \, 2, \, 3)$ \,
in the vacuum Einstein equation.
For example, zero point energy density is given by:
\begin{equation} 
  \int {{d^{4} p}\over (2 \pi)^{4}} \left( \sqrt{m_{B}^{2} + {p}^{2}} 
    - \sqrt{m_{F}^{2} + p^{2}} \right) \, \raisebox{-3mm}{$.$}  
  \label{eq:3.05}
\end{equation}

The five dimensional part may become the sum, rather than the integral, but 
it is still five dimensional vacuum energy and it will contribute to
the five dimensional cosmological constant.

The interaction energy is the correction to the zero point 
energy in the perturbative calculation.
It also will contribute to five dimensional vacuum energy.

There is a different kind of vacuum energy which is of
the four dimensional nature.
The instanton contribution will be of this kind, if it exists.
We explore this possibility later in this section.
\\

We now start from the low energy effective theory in string theory, 
i.e. the pure $E_8$ gauge theory.
The adjoint representation of $E_8$ can be branched into the 
$SU(3) \otimes E(6)$ representations in the following way~\cite{Slansky}: 
\begin{eqnarray} 
  R_{248} = (1, \, 78) + (8, \, 1) 
    + (3, \, 27) + (\overline{3}, \, \overline{27}) \, . 
  \label{eq:3.1}
\end{eqnarray}
We assume that $(1, \, 78)$ and $(8, \, 1)$ remain
as five dimensional gauge fields of $E(6)$ and $SU(3)$
and lie on the line $AOA^{\prime}$.
On the other hand, $(3, \, 27)$ and $(\overline{3}, \, \overline{27})$ 
are assumed to turn into chiral multiplets, although the mechanism 
is not clear.
They also lie on $AOA^{\prime}$.
There is nothing but supergravity in the bulk.
This means that the bulk vacuum solution is provided by the $\lambda = 0$
solution of the previous section.
We, therefore, have the massless gravitino in our model.
Our internal space may stay the same size or may be 
shrinking.
We briefly discuss this issue later in this paper, but it requires 
further investigation.

Each $27$ corresponds to a generation of quarks and leptons of sixteen 
dimensional representation of $SO(10)$, one generation 
of Higgs in the ten dimensional representation and one singlet~\cite{Georgi_Glashow}.
We also have the three generation of quark-lepton like and Higgs like 
objects in $\overline{27}$ representations.
\\

In general, vacuum energy is composed of (1) classical (minimum of
potential) energy which is zero in our case, due to supersymmetry, 
(2) the usual zero point energy $\int d^{n} p \, (E_{B}-E_{F})$, and 
(3) the interaction energies which are created by the supersymmetry 
breaking.
These three vacuum energies will give a 
$\lambda \, g_{\mu \nu} \, (\mu, \, \nu = 0, \, 1, \, 2, \, 3, \, 5)$
term to the five dimensional Einstein vacuum equation and the four 
dimensional part will remain flat if $\lambda$ \lt \, $0$.
However, we also expect the existence of four dimensional vacuum energy
which gives $\lambda \, g_{\mu \nu} \, (\mu, \, \nu = 0, \, 1, \, 2, \, 3)$
if the energy is purely of four dimensional nature.
An example may be provided by the usual instanton.
\\

Let us expand the generic five dimensional field into normal modes:
\begin{equation} 
  \phi_{s} (x_{\mu}, \, \xi) 
    = \textstyle\sum\limits_{n = 0}^{\infty} a_{n} \cos \left( \frac{\ds \pi \xi}{\ds L} n \right) 
  \label{eq:3.2} 
\end{equation}
or
\begin{equation} 
  \phi_{a} (x_{\mu}, \, \xi) 
    = \textstyle\sum\limits_{n = 1}^{\infty} b_{n} \sin \left( \frac{\ds \pi \xi}{\ds L} n \right)
  \label{eq:3.3}
\end{equation}
with $\mu = 0, \, 1, \, 2, \, 3$.
\\

Here, the point $A$ corresponds to $\xi = L$ and the point $A^{\prime}$ 
to $\xi = -L$.
Only $\phi_{S}$ has the massless mode.
We derive an effective four dimensional theory by retaining only the 
massless mode.

This theory will be the $SU(3) \otimes E(6)$ gauge theory with $(3, \, 27)$ 
and $(\overline{3}, \, \overline{27})$ chiral multiplets.
\\

The questions to be asked regarding this four dimensional theory are:
\begin{itemize}
\item[1.] Is it possible to generate a superpotential from the gauge interaction?
\item[2.] Is supersymmetry broken? 
\end{itemize}

First of all, we notice that the most general form of the possible superpotential 
derived from gauge interations takes the following form:
\begin{gather} 
  S = C \int d^{2} \theta \, f (H \overline{H}, \, A) 
  \label{eq:3.4} \\ 
  \intertext{where} 
  H = \varepsilon_{ijk} \, c \, (\mu, \, \nu, \, \rho) \, Q_{\mu}^{i} Q_{\nu}^{j} Q_{\rho}^{k} \, ,
  \label{eq.3.5} \\ 
  \overline{H} 
    = \varepsilon^{ijk} \, \overline{c} \, (\mu, \, \nu, \, \rho) \, \overline{Q}^{\mu}_{i} \overline{Q}^{\nu}_{j} \overline{Q}^{\rho}_{k} \, ,
  \label{eq:3.6} \\ 
  \intertext{and}
  A = Q_{\mu}^{i} \overline{Q}_{i}^{\mu} \, .
  \label{eq:3.7} 
\end{gather}

Here $Q_{\mu}^{i}$ stands for $(3, \, 27)$ quark-lepton-Higgs~\cite{Georgi_Glashow} 
with $i$ for $SU(3)$ triplet and $\mu$ for $E(6)$ $27$-plet.
$c \,(\mu, \, \nu, \, \rho)$ is $27 \times 27 \rightarrow \ol{27}$
Clebsch-Gordon coefficient.
$\overline{Q}_{i}^{\mu}$ is the $(\overline{3}, \, \overline{27})$
quark-lepton-Higgs.
There exists anormaly free $U_{V}(1)$ symmetry where $Q$ and 
$\overline{Q}$ transform with the opposite phases.
This allows only $H \overline{H}$ as an invariant.
\\

We notice that $H$ reduces to the form:
\begin{eqnarray} 
  \varepsilon_{ijk} \, Q_{16}^{i} Q_{16}^{j} Q_{10}^{k}
  \label{eq:3.8}
\end{eqnarray}
in the $SO(10)$ notation.
This is nothing but the usual lepton-quark-Higgs coupling in $SO(10)$.
\\

Is it possible to restrict further our formula for superpotential?
To study this we follow the work of I. Affleck, M. Dine 
and N. Seiberg~\cite{Affleck_Dine_Seiberg} to see if in this case 
we have non-anomalous R-symmetry.
\\

The Lagrangean of our model is 
\begin{eqnarray} 
  \cal{L} 
  &=& {1 \over g^{2}} \Big[ \int d^2 \theta d^2 \overline{\theta} \, \left( Q_{(3, 27)}^{\dag} \, \exp (V) \, Q_{(3, 27)} 
    + \overline{Q}_{(\overline{3}, \overline{27})} \, \exp (-V) \, \overline{Q}_{(\overline{3}, \overline{27})}^{\dag} \right) \nonumber \\
  &+& {1 \over 4} \int d^{2} \theta \, (W_{SU(3)} W_{SU(3)} + W_{E(6)} W_{E(6)}) \nonumber \\
  &+& {1 \over 4} \int d^{2} \overline{\theta} \, (\overline{W}_{SU(3)} \ol{W}_{SU(3)} 
    + \overline{W}_{E(6)} \overline{W}_{E(6)}) \Big] \, 
  \label{eq:3.8}
\end{eqnarray}
in the notation of Wess and Bagger~\cite{Wess_Bagger}.

For simplicity we assumed here a common coupling $g^{2}$ for $E(6)$
and $SU(3)$ at the Planck scale, but it does not have 
to be precise, as long as they are close to each other. 
We also note that due to the existence of $(\overline{3}, \, \overline{27})$
we have a trivial flat direction similar to the case considered 
in~\cite{Affleck_Dine_Seiberg}.
Phenomenologically, we must keep $SU(3)^{c} \times U(1)^{em}$
unbroken in the flat direction, but otherwise it is arbitrary.
\\

Let us consider the following most general form of R-symmetry:
\begin{gather} 
  W_{SU(3)} \, (\theta) 
    = e^{- i \alpha} \, W_{SU(3)} \, (\theta e^{i \alpha}) \, ,  
  \label{eq:3.9} \\
  W_{E(6)} \, (\theta) 
    = e^{- i \alpha} \, W_{E(6)} \, (\theta e^{i \alpha}) \, ,  
  \label{eq:3.10} \\
  Q \, (\theta) 
    = e^{i \beta} \, Q \, (\theta e^{i \alpha}) \, ,  
  \label{eq:3.11} \\
  \intertext{and}
  \overline{Q} \, (\theta) 
    = e^{i \beta} \, \ol{Q} \, (\theta e^{i \alpha}) \, .  
  \label{eq:3.12}
\end{gather}

From the triangular diagram which contains $E(6)$ gauge bosons 
as two external legs we get 
\begin{equation} 
  \alpha +3 \beta = 0 \, . 
  \label{eq:3.13}
\end{equation}
Here we used the Dynkin index~\cite{Slansky} for 
\begin{equation*} 
  E(6): \; \quad \raisebox{1mm}{$\mu$}{}_{78} = 24 
  \quad \mbox{and} \quad
  \raisebox{1mm}{$\mu$}{}_{27} = 6 \, .
  \label{eq:3.14}
\end{equation*}
From the diagram with $SU(3)$ legs we get
\begin{equation} 
  8 \alpha + 9 \beta = 0 \, . 
  \label{eq:3.15}
\end{equation}

The equations \eqref{eq:3.13} and \eqref{eq:3.15} are not compatible 
unless $\alpha = \beta = 0$.
There seems to be no way to restrict the form of $S$ (equation 
\eqref{eq:3.4}) further.
\\

We consider the physics of Planck scale and contemplate what kind of 
dynamics will produce the superpotential of the form \eqref{eq:3.4}.
Although it is not clear at this stage whether the instanton contribution
can give rise to the form \eqref{eq:3.4}, it is worthwhile considering 
the consequences that we get, in case it does.  
\\

We note that we have two kinds of instantons in our model: $E(6)$ 
and $SU(3)$.
Each instanton will lead to the superpotential of the form:
\begin{equation} 
  S = c \, \Lambda^{3 - 6a - 2b} \int d^{2} \theta \, (H \overline{H})^{a} \, A^{b} \, , 
  \label{eq:3.16}
\end{equation}  
where $\Lambda$ is a dynamically generated mass and $c$ is a dimensionless 
constant.
The mass term from the single instanton contribution has the form:
\begin{equation} 
  m = c^{\prime} v \, \exp \, ({-8 \pi} / {g^{2} \left( v \right)}) \, , 
  \label{eq:3.17}
\end{equation}
which should be identified with
\begin{equation} 
  m = cv \left( \frac{\ds v}{\ds \Lambda} \right)^{-3 + 6a + 2b} \, ,
  \label{eq:3.18}
\end{equation}
from equation \eqref{eq:3.16}.
Here $v$ stands for an averaged value of the vacuum value of the scalar 
component of $Q_{\mu}^{i}$ i.e. $\langle A_{\mu}^{i} \rangle$.

The general formula for the running coupling reads:
\begin{equation} 
  \frac{\ds 1}{g^{2} (v)} 
    = \frac{1}{g^{2} (\Lambda)} + \frac{\ds 1}{16 \pi} \, [3 \raisebox{1mm}{$\mu$}_{\lambda} 
    - \Sigma \raisebox{1mm}{$\mu$}_{R} N_{R}] \, \log \frac{v}{\Lambda} \, 
  \label{eq:3.19}
\end{equation}
where $\mu_{\lambda}$ and $\mu_{R}$ stand for Dynkin indices for the adjoint 
representation and for the representation $R$ of the chiral multiplet with 
$N_{R}$ generations, respectively.

From this and equation \eqref{eq:3.18} we get
\begin{equation} 
  m = c^{\prime} v \exp \left( {- 8 \pi^{2}} / {g^{2} (\Lambda)} \right) 
    \left( \frac{\ds v}{\Lambda} \right)^{- {{1}/{2}} (3 \mu_{\lambda} 
    - \Sigma \mu_{R} N_{R})} \, \raisebox{-2mm}{$.$}
  \label{eq:3.20}
\end{equation}
Substituting the Dynkin indices and $N_{R}$ for $E(6)$ or $SU(3)$, we get
\begin{equation} 
  m_{6} = c_{6}^{\prime} v \exp \left( {- 8 \pi^{2}}/{g_{6}^{2} (\Lambda)} \right) 
    \left( \frac{\ds v}{\ds \Lambda} \right)^{-18} \, 
  \label{eq:3.21} 
\end{equation}
for $E(6)$, and
\begin{equation} 
  m_{3} 
    = c_{3}^{\prime} v \exp \left( {- 8 \pi ^{2}}/{g_{3}^{2} \left( \Lambda \right)} \right) 
    \left( \frac{\ds v}{\ds \Lambda} \right)^{+ 18} \, 
  \label{eq:3.22}
\end{equation}             
for $SU(3)$.

$g_{6}(\Lambda)$ stands for the $E(6)$ coupling and $g_{3} (\Lambda)$ 
for the $SU(3)$ coupling.
The sign of the power of $\left( {v}/{\Lambda} \right)$ is the consequence 
of $E(6)$ theory being asymptotic free but $SU(3)$ being infrared free.
\\

The potential which will be derived from our superpotential has two terms,
one from $E(6)$ and one from $SU(3)$:
\begin{eqnarray} 
  V 
  &\cong& (m_{6}^{2} + m_{3}^{2}) v^{2} \nonumber \\ 
  &=& \Lambda^{4} \left[ \exp \left( {- 16 \pi^{2}}/{g_{6}^{2} (\Lambda)} \right) 
    \left( \frac{\ds v}{\ds \Lambda} \right)^{-32} 
    + \omega \exp \left( {-16 \pi^{2}}/{g_{3}^{2} \left( \Lambda \right)} \right) 
    \left( \frac{\ds v}{\ds \Lambda} \right)^{40} \right] \, 
  \label{eq:3.23}
\end{eqnarray}
where $\omega$ is a numerical constant. 
    
The most amusing property of this form is that it has a minimum at
\begin{equation} 
  {v \over \Lambda} 
    = \left( {32 \over 40 \omega} \right)^{1/72} \exp \left\{ {2 \pi^2 \over 9} \left( {1 \over {g_{3}^{2} (\Lambda)}} - {1 \over {g_{6}^{2} (\Lambda)}} \right) \right\} \, \raisebox{-3mm}{$.$}   
  \label{eq:3.24}
\end{equation}

More precisely, we have $V = V \, (v_{\mu}^{i})$ and
\begin{equation} 
{{\partial V} \over {\partial v_{\mu}^{i}}} = 0 \, 
\label{eq:3.245}
\end{equation}
which fixes the flat direction.
Therefore, the flat direction is no longer flat, but it is determined in a non-trivial way.
\\

$g_{3}$ and $g_{6}$ have the same value at the Planck scale ($E_{8}$ 
unifidcation).
We, therefore, have
\begin{equation*} 
  \raisebox{1mm}{$g$}{}_{3}^{2} (\Lambda)
  \quad \mbox{\lt} \, \quad
  \raisebox{1mm}{$g$}{}_{6}^{2} (\Lambda)
\end{equation*}   
for $\Lambda$ $\ll$ Planck scale.
The right hand side will be very large and we expect $v$ to be of the order 
of Planck scale.   
\\

The value of the 'cosmological constant' is, therefore, given by
\begin{eqnarray} 
  \lambda^{4} 
  &\cong& v^{4} \exp \left[ {-16 \pi^{2}}/{g^{2} \left( v \right)} \right] \nonumber \\ 
  &\cong& m_{p}^{4} \exp \left[ {-16 \pi^{2}}/{g^{2} \left( m_{p} \right)} \right] \,  
  \label{eq:3.25}
\end{eqnarray}
where $m_{p}$ is the Planck mass.
From this we have
\begin{eqnarray} 
  \alpha (m_{p}) 
    = \frac{\ds g^{2} (m_{p})}{\ds 4 \pi} 
    = {\pi}/{\log^{m_{p}/{\lambda}}} \, \raisebox{-1mm}{$.$}
  \label{eq:3.26}
\end{eqnarray}
\\

WMAP value~\cite{WMAP_SDSS} of $\lambda \cong 0.01 \, {\rm eV}$ and $m_{p} = 10^{28} \, {\rm eV}$ 
provide us:
\begin{equation*} 
  \alpha \left( m_{p} \right) \cong \frac{\ds 1}{\ds 20.5} \, \raisebox{-3mm}{$.$}
  \label{eq:3.27}
\end{equation*}

The current value for the unification coupling is $\alpha \cong \frac{\ds 1}{\ds 25 \pm 3}$~\cite{OPAL} 
in remarkable agreement with our value, considering the very sensitive 
dependence of $\lambda$ on $g^{2} (m_{p})$.
\\

Let us compare the $SU(N)$ case with $N_{f} = N - 1$, considered in 
reference~\cite{Affleck_Dine_Seiberg} to our case:  We will give special 
attention to the number of fermion zero modes.
\\

\newpage
\no
\underline{$SU(N)$ with $N - 1$ generation of $N$ and $\overline{N}$ quarks}
\\

\begin{itemize}
\item[(1)] number of fermion zero modes when $\langle A_{i} \rangle = 0$
\\

\hspace*{30mm}
\begin{tabular}{|c|c|} \hline
  gaugino & quarks and anti-quarks \\ \hline
  $2N$ & $2 \times \left( N - 1 \right)$ \\ \hline
\end{tabular}
\\

\item[(2)] number of particles which gain mass with $\langle A_{i} \rangle = v_{i}$
\\

\hspace*{30mm}
\begin{tabular}{|c|c|} \hline
  gauge & quarks and anti-quarks \\ \hline
  $N^{2} - 1$ & $N^{2}- 1$ \\ \hline
\end{tabular}
\\

This means that out of all the quarks and antiquarks
\begin{eqnarray*} 
  2 N (N - 1) - (N^{2} - 1) = (N - 1)^{2}
  \label{eq:3.28}
\end{eqnarray*}
quarks and antiquarks remain massless.
\\

\no
Out of $2N$ gaugino zero modes $2 (N-1)$ will couple with the $2 (N-1)$ 
quark-antiquark zero modes.
\\

Finally, out of 2 remaining zero modes, one gets absorbed into the massive quark
and one remains as zero mode which obtains the mass from the instanton.
\\

\end{itemize}

\vspace{\baselineskip}
\no
\underline{$SU(3) \times E_{6}$ case with $(3, \, 27)$ 
and $(\overline{3}, \, \overline{27})$ when $\langle A_{\mu}^{i} \rangle = v_{\mu}^{i} = 0$}
\\

\begin{itemize}
\item[(1)] number of fermion zero modes
\\

\hspace*{30mm}
\begin{tabular}{|c|c|c|} \hline
  $SU(3)$ gaugino & $E(6)$ gaugino & $(3, 27) + (\overline{3}, \overline{27})$ \\ \hline
  $6$ & {} & $2 \times 27$ \\ \hline
  {} & $24$ & $6 \times 3 \times 2$ \\ \hline
\end{tabular}
\\

If we assume $SU(3)^{c} \times U(1)^{em}$ to be maintained in the flat 
direction, we find that the minimum remaining symmetry out of $E(6)$ is $SU(4)$.
The flavor $SU(3)$ is comletely broken.

\no
We get the following:
\\

\item[(2)] number of particles which gain mass with $\langle A_{\mu}^{i} \rangle = \upsilon_{\mu}^{i}$
\\

\hspace*{30mm}
\begin{tabular}{|c|c|c|} \hline
  $SU(3)$ gaugino & $E(6)$ gaugino & $(3, 27) + (\overline{3}, \overline{27})$ \\ \hline
  $8$ & $63$ & $71$ \\ \hline
\end{tabular}
\\

There are $6 \times 27 -71 = 91$ remaining massless quark-lepton-Higgs within
$(3, 27)$ or $(\overline{3}, \overline{27})$.
\\

\no
The question of how many zero modes survive with the constraint of retaining 
$SU(3)^{c} \times U(1)^{em}$ is not clear at this stage. 
There are $2 \times 27 - 6$ remaining zero modes in the $SU(3)$ case and 12 zero
modes in the $E(6)$ case, but we have enough massless and massive modes to absorbe them.
What we need is exactly the one zero mode for each $SU(3)$ or $E(6)$ case which
gains either $m_{3}$ or $m_{6}$ through the instanton.
\\

\no
Whether this is going to happen requires a detailed study which will be given
in a future publication.
\end{itemize}
%
%
\vspace{8mm}
%
%
%
\section{Phenomenology}\label{phenomenology}
\vspace{10mm}

\spc
We discussed the physics of the Planck scale in the previous section.
We would like to consider the low energy physics in this section.
Our model is the $E(6) \times SU(3)$ gauge theory with $(3, 27)$ and 
$(\ol{3}, \ol{27})$ multiplet.
$27$ of $E(6)$ branches into $SO(10)$ representations:
\begin{equation} 
  27 = 1 + 16 + 10 \, .
  \label{eq:4.1}
\end{equation}
We utillize the Georgi-Glashow lepton-quark assignment~\cite{Georgi_Glashow} for 16 and
we assign two $SU(5)$ five dimensional Higgs to 10.
We have three generations of these Higgs multiplets.
Generation $SU(3)$ will be completely broken at the Planck scale, due to the non-zero
$\langle A_{\mu}^{i} \rangle$, but some subgroups of $E(6)$ will remain.

The generated superpotential will have the form
\begin{equation} 
  S = C \int d^{2} \theta \, f \, (H \ol{H}, \, A) \, .
  \label{eq:4.2}
\end{equation}

We assume there is a certain mechanism to produce $S$ even though the 
instanton may not be the one.
We also assume that there is some kind of mechanism to fix the value of 
$\langle A_{\mu}^{i} \rangle = v_{\mu}^{i}$ in the Planck scale 
as in the case of the instanton.

If we expand the expression for $S$ in $1 / v$ we get
\begin{equation} 
  S = \raisebox{1mm}{$\mu$} A +\raisebox{1mm}{$g$} H 
    + \ol{\raisebox{1mm}{$g$}} \ol{H} + O \left({1 \over v} \right) \, \raisebox{-1mm}{$,$}
  \label{eq:4.3}
\end{equation}
where $O \left({1 \over v} \right)$ terms are all non-renormalizable.

$\mu$ is proportional to $v$ ($ve^{- {8\pi / g^{2}}}$
in the instanton case) and $g$, and $\ol{g}$ are dimensionless constants
(they are $ce^{-{8 \pi^{2} / g^{2}}}$ in the instanton case).

$g$ is the running coupling constant and it must be extremely small
because $v_{\mu}^{i}$ is supposed to be in the Planck scale.
$g$ is $ce^{-{8 \pi^{2} / g^{2} (\Lambda)}}$ in the instanton 
case, but $m_{3}$ and $m_{6}$ (which will appear in the four dimensional 
zero point energy) should not run.

$\mu A$ term gives the mixing of $Q$ and $\ol{Q}$.
If $\ol{Q}$ is made high by the boundary condition 
mechanism~\cite{boundary_condition_mechanism}, this gives a small
correction to the $Q$ mass.

The $gH$ term will give the quark-lepton mass directly in terms of Planck mass.
We have
\begin{equation} 
  S_{H} = \raisebox{1mm}{$g$} H
    = \raisebox{1mm}{$g$} \, \varepsilon_{ijk} \, c \, (\mu, \, \nu, \, \rho) \, Q_{16 \mu}^{i} Q_{16 \nu}^{j} Q_{10 \rho}^{k} \, 
  \label{eq:4.4}
\end{equation}
where $c \, (\mu, \, \nu, \, \rho)$ is the $16 \otimes 16 \otimes 10$ 
Clebsh-Gordon coefficient.

Our quark-lepton mass matrix is:
\begin{equation} 
  M^{(i, \mu)(j, \nu)} = \raisebox{1mm}{$g$} \, \varepsilon_{ijk} \, c \, (\mu, \nu, \rho) \, v_{10 \rho}^{k} \, 
  \label{eq:4.5}
\end{equation}
where $v_{10 \rho}^{k} = \langle A_{10\rho}^{k} \rangle$ with 
$A_{10 \rho}^{k}$ the scalar component of $Q_{10 \rho}^{k}$.
\\

We can also check that the right handed neutrino becomes massive directly
by coupling to the gauginos in the flat direction when $v$'s conserve
$SU(3)^{c} \otimes U(1)^{em}$.
We expect all the $v_{\mu}^{i}$ are determined by the equation \eqref{eq:3.245}
or a similar one.
\\

It will be interesting to compare equation \eqref{eq:4.5} with experimental results, 
even leaving the value of $v_{\mu}^{i}$ unfixed.
Similarly, we have
\begin{equation} 
  M_{k}^{(i, \, \mu)(j, \, \nu)} = \raisebox{1mm}{$g$} \, \varepsilon_{ijk} \, c \, (1; \mu, \nu) \, v_{1}^{k} \, ,
  \label{eq:4.6}
\end{equation}
for the three generations of Higgs.
Here, $\mu$ or $\nu$ stands for the ten dimensional component of $SO(10)$.  
$v_{1}^{k}$ is the vacuum value of the $SO(10)$ singlet scalar component.
\\

It is probably worthwhile to compare our model with the usual standard or 
minimal supersymmetric standard models.
\\

The gauge part is more or less the same after the appropriate symmetry breaking.
In the standard model minimization of the potential is done by adding the 
$\varphi^{4}$ term in the Lagrangean, adjusting the mass scale. 
In our case the vacuum values are given by fixing the flat direction 
through equation \eqref{eq:3.245}.
We expect this to be $\simeq 10^{19} \, {\rm GeV}$.
Low mass ($\le {\rm TeV}$) is achieved by extremely small $QQQ$ coupling
which may run as $e^{-{8\pi^{2} / g^{2} (\Lambda)}}$.
Of course, the running of the coupling should flatten out in the low energy 
region because the particles' masses are not very scale dependent experimentally.

The minimum supersymmetric model usually allows maximum number of superpotentials
which are consistent with renormalizability.
On the other hand, we assume the potentials must be produced by the gauge 
interaction through a certain mechanism such as the instanton.

Our renormalizable superpotential terms are $H$, $\ol{H}$ and $A$, simply 
due to the symmetry argument.
Among these, $\ol{H}$ and $A$ may not play important roles in the low energy 
region if $(\ol{3}, \, \ol{27})$ are all high.
All the non-renormalizable superpotential terms are suppressed by the inverse 
power of $v$.

We are, therefore, left with the single term $H$ in the low energy.
The existence of a flat direction and its fixation is necessary in our model.

\subsubsection*{\underline{Supersymmetry and gauge symmetry breaking}} 

\vspace{3mm}
\spc

The fact that the fifth-dimension is compact can be used to break either 
supersymmetry (Sherk-Shwartz) or gauge symmetry (a boundary condition).

\enlargethispage{-5mm}

In our gauge model bosonic physical quantities must be periodic, but they 
are quadratic or higher in the field variables.
This enables us to have the fields either periodic (equation \eqref{eq:3.2}) 
or anti-periodic (equation \eqref{eq:3.3}).
Even a part of super or gauge multiplet can be different 
from the other part.

There is no physical quatity which transforms as $SU(3) \, {\rm triplet} \otimes \, SU(2)$
doublet, thus enabling the triplet Higgs to have a large mass.
Quark-lepton superpartners can be treated similarly.
\\

The only condition for the case of supersymmetry is $\lambda_{5}$ \lt \, $0$.
If the zero point energy dominates $\lambda_{5}$, then the fermion part 
must be heavier than the boson part in total.
Still, there will be more than one solution which can achieve this case.

If we can measure $\lambda_{5}$, it will give a stringent restriction on 
the possible solutions.
\\

Phenomenologically, one of the most significant consequences is that the 
superpartners in the quark-lepton-Higgs sector must be as heavy as the 
triplet Higgs, since these are all of the order of $1 / L$ where $L$ is 
the size of the internal space.
The triplet Higgs must be so heavy as to not allow the proton decay to 
occur frequently.

This indicates that there is no chance we will find superparticles either 
by the LHC or by the ILC.
It will be exciting, though, to find three generations of Higgs particles 
in these machines.
The existence of $(\ol{3}, \, \ol{27})$ in the low energy region is not 
totally excluded either.
\\

There are two paths of gauge symmetry breaking: 
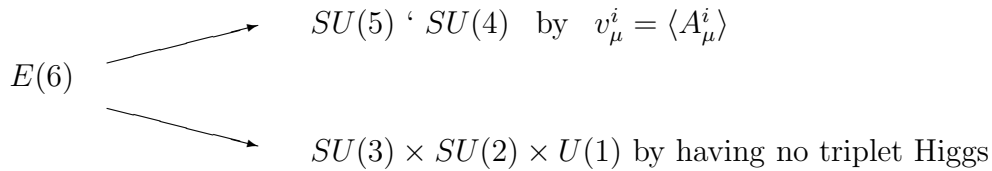
\begin{figure}[h] 
  \begin{center}
    \setlength{\unitlength}{1mm}
    \begin{picture}(200,25)(0,0) 
      \put(15, 10){$E(6)$}
      \put(28,13){\vector(4,1){20}}
      \put(28,7){\vector(4,-1){20}}
      \put(55,17){$SU(5)$ ` $SU(4)$ \, by \, $v_{\mu}^{i} = \langle A_{\mu}^{i} \rangle$} 
      \put(55,0){$SU(3) \times SU(2) \times U(1)$ by having no triplet Higgs}
    \end{picture}
  \end{center}
\end{figure}
\\

The low energy physics beyond TeV region will be a competition between 
$SU(3) \times SU(2) \times U(1)$ and $SU(4)$  if we have such 
$\langle v_{\mu}^{i} \rangle$ which violates $SU(5)$ down to $SU(4)$.
%
%
\vspace{15mm}
%
%
%
\section{Concluding Remarks}\label{conclusion}
\vspace{10mm}

\spc
The main achievement of this work is finding solutions 
to the $n$-dimensional Einstein equation in which a certain 
extra-dimensional direction is chosen which plays the role of Euclidean time.
We find that the warp solution of Randall and Sundrum~\cite{Randall_Sundrum} 
(without the $\delta$ function source) is dual to the expanding 
universe solution with Euclidean time and Minkowski time 
playing the role of dual coordinates.

We also consider the case of $\lambda = 0$ which admits a solution 
of expanding vacuum with a shrinking internal space.
In our approach the geometry of the vacuum is determined by the cosmological constant
which in turn is determined by the vacuum energy.
\\

Matter must exist in a self-consistent way with geometry.
Therefore, we propose a model in which matter (including the 
gauge field) lies on five dimensional spacetime with the extra 
Euclidean time.
This enables most of the vacuum energy (including the zero-point energy 
and its gauge corrections) to be of the five dimensional nature.
This leaves the four dimensional spacetime to be flat as the Randall-Sundrum
warp solution suggests.
Supersymmetry can be broken, as long as $\lambda_{5}$ \lt \, $0$.
\\

An intering property of $n$-dimensional vacuum Einstein equation with 
$\lambda = 0$ is the existence of a solution in which the usual three 
dimensional space is expanding, but the interval space is shrinking.

If we adopt this solution, we may claim that our three dimensional space 
started from the Planck size, but there emerges some mystery about the 
internal space.
If it also starts from the Planck size, the current internal space size 
will be so small that it is essentially zero.

If we want internal space size to be somewhat larger than the Planck 
length, so that we obtain an experimentally reachable proton lifetime, 
then we have to explain why the internal space started from a larger size.
The only acceptable solution for the internal space may be that it stays 
at the Planck scale.
This point should be investigated further.
\\

Our proposed model is based on~\cite{Slansky}:
\begin{equation*}
  E(8) \longrightarrow SU(3) \times E(6) \, .
\end{equation*}
The entire matter is contained in the adjoint representation of $E(8)$.
Among these, we assume $(3, \, 27)$ and $(\ol{3}, \, \ol{27})$ (in the
$SU(3) \times E(6)$ notation), turn into chiral multiplets.

Matter lies on the Euclidean time $\otimes$ (our spacetime).
We do not allow any superpotential, unless it is generated by gauge 
interactions.

An attractive candidate is the instanton contribution, altough other
possibilities cannot be excluded.
The advantage of the instanton contribution is that it gives a four
dimensional vacuum energy by violating supersymmetry in a tiny amount.
\\

The quark-lepton-Higgs masses are determined by fixing the flat 
direction in the sense of Affleck, Dine and Seiberg~\cite{Affleck_Dine_Seiberg}.
We expect that the instanton contribution (or something similar) will 
establish that the flat direction parameters will be in the Planck scale,
and thus, the quark-lepton-Higgs masses will be determined directly by 
the Planck mass.
\\

We have not made the distinction between the Planck scale and the grand
unified scale~\cite{Pati_Salam_et_al} in this paper.
In fact, for example, the expression for the cosmological constant 
$v^{4} \exp (-{16 \pi^{2} / g^{2} (v)})$ gives more or less 
the same value for $g^{2} (v)$ whether we take $v$ to be of 
the Planck scale or the grand unified scale.
$g^{2} (v)$ is only logarithmically dependent on the scale.
It is a matter of detail in our model to distinguish these two scales.
\\
%
%
\vspace{20mm}
%
%
%
\section*{Acknowledgments}
\vspace{10mm}
\spc

I would like to thank Y. Shimizu for checking the solutions to 
the Einstein equation using the ''Reduce'' program and S. Traweek 
for checking my English.

This work could not have been done without the encouragement of my father 
(who passed away February 3, 2006) who insisted that the author (sixty-eight 
years old) is now at the highest point of his research carrier.
\vspace{20mm}
%

%
%

%
%
%
%

\end{document}